\documentclass[aps,showpacs,superscriptaddress,a4paper,nofootinbib]{revtex4}
\pdfoutput=1

\usepackage{graphicx, color}
\usepackage{epstopdf}           % must come after graphicx

\usepackage{amsmath}
\usepackage{amssymb}
\usepackage{color}
\usepackage{graphicx}
\usepackage{latexsym}
\usepackage{times}

\newcommand{\scri}{\ensuremath{\mathcal{J}^+}}

\def\be{\begin{equation}}
\def\ee{\end{equation}}
\def\bea{\begin{eqnarray}}
\def\eea{\end{eqnarray}}
\def\scri{{\mathcal I}^+}

\begin{document}

\title{The gravitational wave strain in the
characteristic formalism of numerical relativity}

\author{Nigel T. Bishop}
\affiliation{
  Department of Mathematics,
  Rhodes University,
  Grahamstown, 6139
  South Africa
}
\affiliation{
Faculty of Engineering and Natural Sciences, Sabanci University, Orhanli - Tuzla 34956,
Istanbul, Turkey}

\author{Christian Reisswig}
\thanks{Einstein Fellow}
\affiliation{
  Theoretical Astrophysics Including Relativity,
  California Institute of Technology,
  Pasadena, CA 91125, USA
}

\begin{abstract}
The extraction of the gravitational wave signal, within the context of a characteristic numerical
evolution is revisited. 
A formula for the gravitational wave strain is developed and tested, and is made publicly 
available as part of the PITT code within the Einstein Toolkit.
Using the new strain formula, we show that artificial non-linear 
drifts inherent in time integrated waveforms 
can be reduced for the case of a binary black hole merger configuration. 
For the test case of a rapidly spinning stellar core collapse model,
however, we find that the drift must have different roots.
\end{abstract}


\maketitle
\section{Introduction}

The characteristic approach to numerical relativity has been developed over a number of years
(see, for example, the review~\cite{Winicour05}), and the method is often used, as
characteristic extraction~\cite{Reisswig:2009us,Reisswig:2010a, Babiuc:2011b, Babiuc:2011}, in the computation of
waveforms from astrophysical events (see e.g.,~\cite{Reisswig:2013sqa, Ott:2011, Pollney:2010hs, Hinder:2013oqa}). 
The origin of the approach is the work of Bondi
{\it et al.}~\cite{Bondi62} in which coordinates particularly suitable for the description
of gravitational radiation were introduced, and the corresponding Einstein field equations
were derived. It was also shown that it is always possible to impose additional conditions on the
coordinates, now called the Bondi gauge, so as to greatly simplify the asymptotic form of
the metric. In this gauge, the description of gravitational radiation in terms of the metric,
is very simple. In addition, Bondi gauge is the coordinate frame 
in which a gravitational wave detector would measure gravitational radiation.

One difficulty faced by characteristic numerical relativity is that, in general, it is not
possible to impose the Bondi gauge conditions since they are asymptotic conditions on the
coordinates. In practice, the gauge is fixed in the interior of the spacetime.
For example, in characteristic extraction \cite{Winicour05, Reisswig:2010a, Babiuc:2011} the 
coordinates and therefore the gauge are induced by the gauge driver of the 3+1 Cauchy evolution system at the
extraction worldtube. It is non-trivial to derive an expression from metric data
in a general gauge that describes the gravitational radiation field as measured by an
asymptotic inertial observer. This was done for the gravitational news ${\mathcal N}$
at the same time as the development of the first 3D characteristic code, now known as the
PITT code~\cite{Bishop97b}, and it was only quite recently, in 2009~\cite{Babiuc:2009}, that a
formula for $\psi_4$ (which is commonly used to describe gravitational radiation in numerical
relativity) was developed. It should also be mentioned that there is an alternative approach
to calculating ${\mathcal N}$
based on an explicit construction of the coordinate transformation between the Bondi and
general gauges~\cite{Bishop03}, but the method has not been further developed or implemented.

While formulas have been developed for ${\mathcal N}$ and $\psi_4$ within characteristic 
numerical relativity, there is no specific formula in the literature for the gravitational
wave strain $(h_+,h_\times)$, which is what would be measured by a gravitational wave detector.
Instead, the strain is found by time integration of ${\mathcal N}$ or $\psi_4$, and it appears
that this process introduces aliasing or other numerical effects so reducing the accuracy of the
result~\cite{Reisswig:2011, Reisswig:2011a, Baker:2002qf, Berti:2007snb}.

This work re-investigates gravitational radiation within the characteristic
formalism, and derives a formula for the wave strain. In addition, alternative formulas for
some intermediate variables are found that avoid time integration. In this way it is possible
to reduce, but not to eliminate, the number of time integrals used in the computation of
gravitational radiation descriptors.

Sec.~\ref{s-rev} summarizes previous relevant work, and then Sec.~\ref{s-coord} introduces
and evaluates the properties of the coordinate transformation
between the Bondi and general gauges. Sec.~\ref{s-strain} derives a formula for the gravitational wave
strain, and Sec.~\ref{s-form} finds an explicit formula (rather than an evolution equation) for
the phase of ${\mathcal N}$ and $\psi_4$. Numerical tests, and other approaches to validation,
are discussed in Sec.~\ref{s-num}. Sec.~\ref{s-con}
is the Conclusion. This work makes extensive use of computer algebra scripts; details are
given in an Appendix, and the scripts are available in the online supplement.

\section{Review of previous work}
\label{s-rev}

\subsection{The characteristic formalism}
The formalism for expressing Einstein's equations as an evolution system based
on characteristic, or null-cone, coordinates is based on work originally due to Bondi
{\it et al.}~\cite{Bondi1960,Bondi62} for axisymmetry, and extended to the general case
by Sachs~\cite{Sachs62}. The formalism is covered
in the review by Winicour~\cite{Winicour05}. We start with coordinates based upon a
family of outgoing null hypersurfaces.
Let $u$ label these hypersurfaces, $x^A$ $(A=2,3)$ label
the null rays, and $r$ be a surface area coordinate. In the resulting
$x^\alpha=(u,r,x^A)$ coordinates, the metric takes the Bondi-Sachs
form
\be
 ds^2  =  -\left(e^{2\beta}(1 + W_c r) -r^2h_{AB}U^AU^B\right)du^2
        - 2e^{2\beta}dudr -2r^2 h_{AB}U^Bdudx^A 
        +  r^2h_{AB}dx^Adx^B,
\label{eq:bmet}
\ee
where $h^{AB}h_{BC}=\delta^A_C$ and
$\det(h_{AB})=\det(q_{AB})$, with $q_{AB}$ a metric representing a unit
2-sphere; for the computer algebra calculations presented later it is necessary
to have specific angular coordinates, and for that purpose stereographic
coordinates are used with $x^A=(q,p)$ and
\be
q_{AB}dx^A dx^B=\frac{4}{(1+q^2+p^2)^2}\left(dq^2+dp^2\right).
\ee
$W_c$ is a normalized variable used in the code, related to the usual Bondi-Sachs
variable $V$ by $V=r+W_c r^2$. It should be noted here that different
references use various notations for what is here denoted as $W_c$, and in
particular ref.~\cite{Bishop97b} uses $W$ with $W=r^2W_c$. It is convenient to
describe angular quantities and derivatives by means of complex numbers using the
spin-weighted formalism and the eth ($\eth$) calculus~\cite{Bishop96,Bishop97b,Gomez97}.
To this end, $q_{AB}$ is represented by a complex dyad $q_A$
with, for example, $q_A=(1,i)2/(1+q^2+p^2)$ in stereographic coordinates. Then
$h_{AB}$ can be represented by its dyad component $J=h_{AB}q^Aq^B/2$.
We also introduce the fields $K=\sqrt{1+J \bar{J}}$ and $U=U^Aq_A$.

We will be concerned with asymptotic properties, and so make the coordinate
transformation $r$ to $\rho=1/r$, and consider the behaviour of quantities near $\rho=0$.
We have
\begin{equation}
   ds^2= \rho^{-2}\left(-\left(e^{2\beta}(\rho^2+\rho W_c)
 -h_{AB}U^AU^B\right)du^2
        +2e^{2\beta}dud\rho -2 h_{AB}U^Bdudx^A + h_{AB}dx^Adx^B\right).
   \label{eq:lmet}
\end{equation}
In contravariant form,
\be
{g}^{11}=e^{-2\beta}\rho^3(\rho+ W_c) , \;\; {g}^{1A}=\rho^2e^{-2\beta} U^A,
{g}^{10}=\rho^2e^{-2\beta}, \;\;
{g}^{AB}=\rho^2h^{AB}, \;\; {g}^{0A}={g}^{00}=0.
\label{e-bsc}
\ee

\subsection{The Bondi gauge}
The spacetimes considered here are asymptotically flat; but even so, in general, the metric
near $\rho=0$ may be complicated. Matters are greatly simplified by using coordin‌ates which
naturally exhibit the property of asymptotic flatness. As shown in the original work of
Bondi {\it et al.}~\cite{Bondi62}, such coordinates always exist, and have become known as the
Bondi gauge. We will use $\tilde{}\;$ to denote a quantity in the Bondi gauge, e.g. the Bondi gauge
coordinates are denoted by $(\tilde{u},\tilde{\rho},\tilde{x}^A)$. The Bondi gauge can be defined
by conditions on the metric quantities
\be
\lim_{\tilde{\rho}=0}(\tilde{J},\tilde{U},\tilde{\beta},\tilde{W}_c)=0;
\ee
since flat Minkowski spacetime has $J,U,\beta,W_c=0$ everywhere, it is clear that the Bondi
gauge gives a natural expression of asymptotic flatness. Actually, we can go further
and use the asymptotic Einstein vacuum equations, that is evaluate the Ricci tensor as a power
series in $\tilde{\rho}$ and keep only the leading order terms. In this way the dependence
of the metric coefficients to leading order on $\tilde{\rho}$ is found to
be~\cite{Tamburino:1966,Isaacson83}
\be
\tilde{J}=0+{\mathcal O}(\tilde{\rho}),\;
\tilde{U}=0+{\mathcal O}(\tilde{\rho}^2),\;
\tilde{\beta}=0+{\mathcal O}(\tilde{\rho}^2),\;
\tilde{W}_c=0+{\mathcal O}(\tilde{\rho}^3).
\ee
It will be convenient to describe gravitational radiation with respect to an orthonormal null
tetrad, so that formulas are independent of specific coordinates. The normalization adopted is
\be
\ell^a\ell_a=n^a n_a=m^a m_a =\ell^a m_a =n^a m_a=0, \ell^a n_a=m^a \bar{m}_a=1,
\ee
which is consistent with common practice in numerical relativity, although Refs.~\cite{Bishop97b,
Winicour05,Babiuc:2009} normalize to 2 rather than to 1.
In the Bondi gauge, the leading order terms of the natural null tetrad are
\begin{equation}
\tilde{\ell}^\alpha=\left(0,- \frac{\tilde{\rho}^2}{\sqrt{2}},0,0\right),\;
\tilde{n}^\alpha=\left(\sqrt{2},\frac{\tilde{\rho}^2}{\sqrt{2}},0,0\right),\;
\tilde{m}^\alpha=\left(0,0,\frac{\tilde{\rho} \tilde{q}^A}{\sqrt{2}}\right).
\label{e-ntMpu}
\end{equation}

\section{Coordinate transformation between the Bondi and general gauges}
\label{s-coord}
We construct quantities in the general gauge by means of a coordinate transformation to
the Bondi gauge. The form of the transformation has been used
previously~\cite{Bishop03},
and is written as a series expansion in $\rho$ with coefficients arbitrary functions of
the other coordinates. Thus it is a general transformation, and the requirements that
$g^{ab}$ must be of Bondi-Sachs form, and that $\tilde{g}^{ab}$ must be in the Bondi
gauge, imposes conditions on the transformation coefficients. The transformation is
\begin{equation}
u \rightarrow {\tilde u}=u+u_0+\rho A^u,
\rho \rightarrow {\tilde \rho}=\rho \omega+\rho^2 A^\rho,
x^A \rightarrow {\tilde x}^A=x^A+x^A_0+\rho A^A,
\label{e-bs2b}
\end{equation}
where $u_0,A^u,\omega,A^\rho,x^A_0,A^A$ are all functions of $u$ and $x^A$ only. We obtain
\begin{equation}
{{\tilde g}}^{ab}=\frac{\partial {\tilde x}^a}{\partial x^c}
\frac{\partial {\tilde x}^b}{\partial x^d} { g}^{cd}\;\mbox{and }
{ g}_{ab}=\frac{\partial {\tilde x}^c}{\partial x^a}
\frac{\partial {\tilde x}^d}{\partial x^b}{\tilde g}_{cd}.
\end{equation}
We find the following formulas (many of which have been given
previously~\cite{Bishop97b,Bishop03})
\be
\tilde{\rho}^2+\mathcal{O}(\tilde{\rho}^3)={{\tilde g}}^{01}, \;\;\mbox{thus }
\tilde{\rho}^2=\omega  e^{-2\beta}(1+u_{0,u}+U^Au_{0,A})\rho^2,\;\;
\mbox{so that }\;(\partial_u+U^B\partial_B)u_{0}=\omega e^{2\beta}-1,
\label{e-du0}
\ee
 \be
\mathcal{O} (\tilde{\rho}^3)={{\tilde g}}^{A1}, \;\;\mbox{thus }
0=\omega e^{-2\beta}(x^A_{0,u}+U^A+U^B x^A_{0,B})\;
 \mbox{so that } (\partial_u+U^B\partial_B)x^A_0=-U^A,
 \label{e-dx0}
\ee
\be
\mathcal{O} (\tilde{\rho}^4)={{\tilde g}}^{11}, \;\;\mbox{thus }
0=\omega e^{-2\beta}(2\omega_{,u}+2U^A \omega_{,A}+\omega W_c)\;
 \mbox{so that } (\partial_u+U^B\partial_B)\omega=-\omega W_c/2,
 \label{e-dom}
 \ee
 \be
 0 ={\tilde g}^{00}\;\; \mbox{so that to }\mathcal{O}(\tilde{\rho}^2),\;\;
 2\omega A^u=\frac{J\bar{\eth}^2u_0+\bar{J}\eth^2 u_0}{2}-K\eth u_0 \bar{\eth}u_0.
 \label{e-omAu}
 \ee
In the next equations, $X_0=q_A x_0^A, A=q_A A^A$; the introduction of
these quantities is a convenience to reduce the number of terms in the
formulas, since $x_0^A, A^A$ do not transform as 2-vectors. As a result, the quantity
$\zeta=q+ip$ also appears, and the formulas are specific to stereographic coordinates.
\begin{eqnarray}
0 &=&\tilde{q}_A {\tilde g}^{0A}\; \;\; \mbox{so that to }\mathcal{O}(\tilde{\rho}^2),
\nonumber \\
0&=&2A\omega+2A_u X_0 U \bar{\zeta}  e^{-2\beta} +K\eth u_0 (2 +\bar{\eth}X_0+2X_0\bar{\zeta})
+K\bar{\eth}u_0\eth X_0 -J \bar{\eth}u_0 (2+\bar{\eth}X_0+2 X_0\bar{\zeta})-\bar{J}\eth u_0 \eth X_0\, .
\label{e-fA}
\end{eqnarray}
The following formulas were derived using the transformation of the covariant metric from
Bondi to general gauge
\be
\det(q_{AB})=\det(g_{AB})\rho^4 \;\;
\mbox{so that }
\omega=\frac{1+q^2+p^2}{1+\tilde{q}^2+\tilde{p}^2}
      \sqrt{1+q_{0,q}+p_{0,p}+q_{0,q}p_{0,p}-q_{0,p}p_{0,q}}\, ,
\label{e-exom}
\ee
\be
J=\frac{q^A q^B g_{AB}}{2} \rho^2\;
\mbox{so that } J=
\frac{(1+q^2+p^2)^2}{2(1+\tilde{q}^2+\tilde{p}^2)^2\omega^2}
\eth X_0 (2+\eth\bar{X}_0+2\bar{X} \zeta).
\label{e-fJ}
\ee

\section{Formulas for gravitational radiation}
\label{s-strain}
Formulas for $\psi_4$ and ${\mathcal N}$ have been derived in earlier
work~\cite{Bishop97b,Babiuc:2009}, and take the form
\be
{\mathcal N}=\frac{e^{-2i\delta} F^a F^b N_{ab}}{2\omega^2},\qquad
\lim_{r\rightarrow \infty} r\psi_4=\lim_{\rho\rightarrow 0}
\frac{e^{2i\delta} \rho {C}_{abcd} n^a \bar{F}^b n^c \bar{F}^d}{\omega^3},
\label{e-Npsi4}
\ee
where $F^a$ is related to the null tetrad vector $m^a$ by
\be
m^a= e^{-i\delta(u,x^A)}\rho F^a
\;\;\mbox{with } \;
F^a= \left(0,0,\frac{q^A\sqrt{K+1}}{2}-\frac{\bar{q}^A J}{2\sqrt{(K+1)}}\right).
\label{e-mF}
\ee
The Weyl tensor ${C}_{abcd}$, the null tetrad vector $n^a$ and the quantity $N_{ab}$
are not specified explicitly here because these details will not be needed. The quantity
$\delta(u,x^A)$ is a phase factor which has been fixed in previous
work~\cite{Bishop97b,Babiuc:2009} by an evolution equation derived from the condition
that the null tetrad vector $m^a$ be parallel propagated along the generators of $\scri$;
an alternative expression for $\delta$, which has the advantage of being explicit, will be
derived below.

A gravitational wave detector does not measure ${\mathcal N}$ or $\psi_4$, but instead
measures the gravitational wave strain $h$ (i.e. $h_+$ and $h_\times$), which in the Bondi gauge
is related to the metric variables used here by
\be
h\equiv \lim_{\tilde{r}\rightarrow\infty}\tilde{r}\left(h_+ +i h_\times\right)
=\tilde{J}_{,\tilde{\rho}}.
\ee
In the Bondi gauge, expressions for ${\mathcal N}$ and $\psi_4$ are particularly simple
\be
{\mathcal N}= \frac{\tilde{J}_{,\tilde{\rho}\tilde{u}}}{2},\;
\lim_{\tilde{r}\rightarrow \infty} \tilde{r}\psi_4
= \bar{\tilde{J}}_{,\tilde{\rho}\tilde{u}\tilde{u}}\; ,
\ee
from which it follows that $h$, ${\mathcal N}$ and $\psi_4$ are simply related
\be
h_{,\tilde{u}\tilde{u}}=2 {\mathcal N}_{,\tilde{u}}=
\left(\lim_{\tilde{r}\rightarrow \infty} \tilde{r}\bar{\psi}_4\right).
\ee
In previous work, $h$ has been found by integrating
${\mathcal N}$ or $\psi_4$ with respect to Bondi gauge time $\tilde{u}$.
In practice, time integration has been observed to introduce aliasing
~\cite{Reisswig:2011, Reisswig:2011a}, that is a non-linear drift of the mean of the wave
oscillation away from zero. Thus,
it is useful to derive a direct formula for $h$.

\subsection{Formula for the gravitational wave strain $h$}
The calculation starts from $\tilde{g}^{ab}$ as a series
in $\tilde{\rho}$ and applies the
coordinate transformation \eqref{e-bs2b} to find $g^{ab}$, now as a series in $\rho$. Then
\be
J=\frac{q_A q_B g^{AB}}{2\rho^2}
\ee
is also expressed as a series, and the second term of the series is $J_{,\rho}$. The
expression found for $J_{,\rho}$ is linear in $\tilde{J}_{,\tilde{\rho}}$, and takes the form
\be
C_1 J_{,\rho}= C_2 \tilde{J}_{,\tilde{\rho}} +C_3 \tilde{\bar{J}}_{,\tilde{\rho}}+C_4,
\ee
which may be inverted to give
\begin{eqnarray}
h=\tilde{J}_{,\tilde{\rho}}&=&\frac{C_1\bar{C}_2 J_{,\rho}-C_3 \bar{C}_1 \bar{J}_{,\rho}
            +C_3 \bar{C}_4 -\bar{C}_2 C_4}{\bar{C}_2 C_2-\bar{C}_3 C_3}\;\;\;
\mbox{where the coefficients are}\nonumber \\
C_1&=&\frac{4\omega^2 (1+\tilde{q}^2+\tilde{p}^2)^2}{(1+q^2+p^2)^2},\;
C_2=\omega (2+\eth\bar{X}_0+2\bar{X}_0\zeta)^2,\;
C_3=\omega (\eth X_0)^2,\nonumber \\
C_4&=&\eth A (4+2\eth \bar{X}_0+4 \bar{X}_0\zeta)+\eth X_0 (2\eth\bar{A}+4\bar{A}\zeta)
+4\eth\omega \eth u_0.
\label{e-Jtrht}
\end{eqnarray}
The above formula for the wave strain involves intermediate variables, and the procedure for
calculating them is to solve equations for the variable indicated in the following order:
Eq.~(\ref{e-dx0}) for $x_0^A$ and thus $X_0$, Eq.~(\ref{e-exom}) for $\omega$,
Eq.~(\ref{e-du0}) for $u_0$, Eq.~(\ref{e-omAu}) for $A^u$, and Eq.~(\ref{e-fA}) for $A$.

The linearized approximation may be used for the Einstein field equations and for the description
of gravitational radiation when $|J|,|U|,|\beta|,|W_c|\ll 1 $. This case is important because
it often applies in astrophysical problems in which characteristic extraction is
used~\cite{Reisswig:2010a}, and formulas for ${\mathcal N}$ and $\psi_4$ have been given
previously~\cite{Bishop-2005b,Babiuc:2009}. In this approximation, the
wave strain Eq.~(\ref{e-Jtrht}) simplifies to
\be
h={J}_{,{\rho}}-\eth A,
\label{e-JtrhtLin}
\ee
and linearization reduces Eq.~(\ref{e-fA}) to $A+\eth u_0=0$ so that
\be
h={J}_{,{\rho}}+\eth^2 u_0\, .
\label{e-Jtrhtl}
\ee

\subsection{Gauge freedom of the wave strain $h$}
From Eq.~(\ref{e-du0}), it is clear that the quantity $u_0$ defined in the coordinate
transformation Eq.~(\ref{e-bs2b}) is subject to the gauge freedom $u_0\rightarrow
u_0^\prime = u_0 + u_G$, provided that $(\partial_u +U^A\partial_A)u_G=0 $; in the
linearized case this is simply $\partial_u u_G=0$ so that $u_G=u_G(x^A)$. Thus the
wave strain $h$ is subject to the gauge freedom $h\rightarrow h^\prime = h + h_G$,
where, in the linearized case,
\be
h_G=\eth^2 u_G (x^A).
\ee
In the general case, the relation between $h_G$ and $u_G$ is via Eqs.~(\ref{e-omAu}),
(\ref{e-fA}) and (\ref{e-Jtrht}) and is much more
complicated, but these details will not be needed. Decomposing $u_G$ into spherical
harmonics, it follows that, in the linearized case, the gauge freedom is represented
by the addition of a constant to each spherical harmonic component of $h$. In the common case
of a numerical simulation in which the final state is the Kerr geometry, these constants would
normally be set so that $h=0$ at the final time.

The same gauge freedom would occur if the wave strain $h$ is obtained by time integration
of the news ${\mathcal N}$, in this case appearing as an arbitrary ``constant'' of integration
$f(\tilde{x}^A)$.

\subsection{Explicit formula for the phase factor $\delta$}
\label{s-form} 
We now revisit the computation of the phase factor $\delta$ introduced in
Eq.~(\ref{e-Npsi4}). We find an expression for $\delta$ which does not require
solution of an ODE, and thus reduces the number of time integrals required for
computing $\psi_4$ or $\mathcal{N}$.
The null tetrad vector $m^a$ in Eq.~(\ref{e-mF}) and $\tilde{m}^a$ in Eq.~(\ref{e-ntMpu})
must be related by the coordinate transformation Eq.~(\ref{e-bs2b}). Thus
\be
\frac{\tilde{q}^A}{\sqrt{2}}=
\frac{\partial x^A_0}{\partial x^B}e^{-i\delta}\frac{F^B}{\omega}
\label{e-m2mt}
\ee
Then $\delta$ is fixed by contracting Eq.~(\ref{e-m2mt}) with $\bar{\tilde{q}}_A/\sqrt{2}$,
since the LHS evaluates to $1$ and the only unknown is $\delta$,
\be
e^{i\delta}=\frac{F^B \bar{\tilde{q}}_A}{\omega\sqrt{2}} \frac{\partial x^A_0}{\partial x^B} 
=\frac{1+q^2+p^2}{4\omega(1+\tilde{q}^2+\tilde{p}^2)}\sqrt{\frac{2}{K+1}}
\left((K+1)(2+\eth \bar{X}_0+2\bar{X}_0\zeta)-J\bar{\eth}\bar{X}_0\right).
\label{e-delta}
\ee

\section{Numerical tests}
\label{s-num}

\subsection{Analytic verification}

At the analytic level there are some simple consistency tests that can be used to check
aspects of the formulas derived here. First, in the linearized case, differentiation
of the LHS of  Eq.~(\ref{e-Jtrhtl}) with respect to $\tilde{u}$ is $2{\mathcal N}$; now in
the linearized case, $\partial_{\tilde{u}}=\partial_u$ so $2{\mathcal N}=J_{,u\rho}+\eth^2 u_{0,u}$,
and then applying Eq.~(\ref{e-du0}) (linearized) gives
\be
2{\mathcal N}=J_{,u\rho}+\eth^2\omega +2\eth^2\beta,
\ee
which is consistent with previous results~\cite{Bishop-2005b,Babiuc:2009}.

The validity of Eq.~(\ref{e-m2mt}) can be checked by contraction with $\tilde{q}_A$, since
the LHS reduces to $0$. Computer algebra was used to replace $J,\omega$ on the RHS 
with terms of the form $x_{0,B}^A$ using Eqs.~(\ref{e-fJ}) and (\ref{e-exom}); the
result was identically $0$.
The RHS of Eq.~(\ref{e-delta}) must be a complex number of modulus $1$. In this case the
computer algebra was unable sufficiently to simplify the expression, so instead random values
for the variables were generated and the resulting formula evaluated numerically. The result for
the absolute value was found to be $1$ (to within computational precision).

\subsection{Tests of the strain formula}

\begin{figure}
  \includegraphics[width=0.49\linewidth]{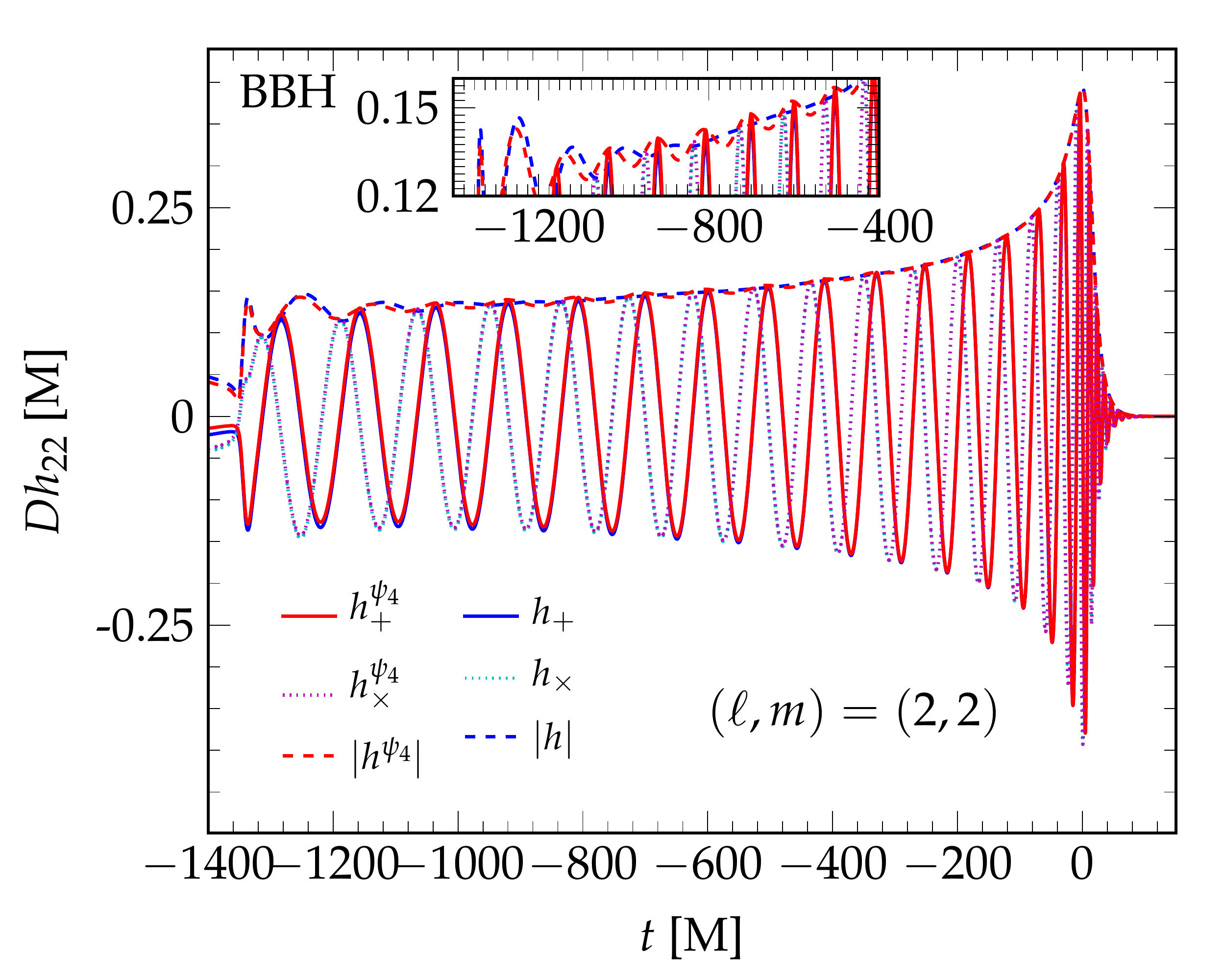}
  \includegraphics[width=0.49\linewidth]{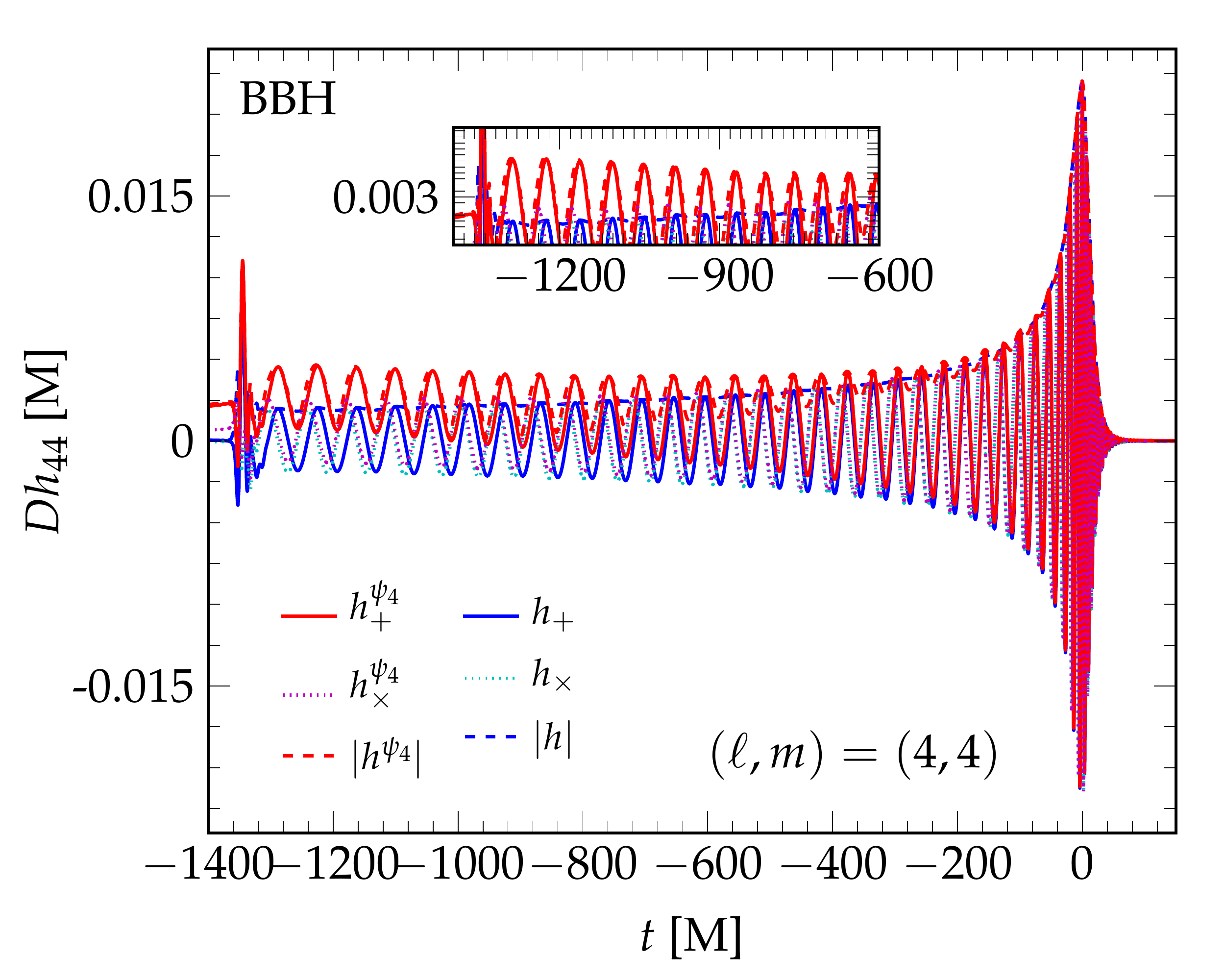}
  \caption{\textit{Left}: The ``+'' and ``$\times$'' polarizations, as well as the amplitude $|h|$ of the $(\ell,m)=(2,2)$ mode of the GW strain 
     emitted from a non-spinning equal-mass binary black hole merger (i) computed using formula \eqref{e-Jtrhtl}, and
     (ii) time integrated from $\psi_4$. The inset shows a close-up of the amplitude evolution. The amplitude of the time-integrated $h^{\psi_4}$ 
     is subject to larger oscillations compared to the amplitude of $h$ computed using
     formula \eqref{e-Jtrhtl}, which however decay during the late inspiral. \textit{Right}: The same as in the left panel, but for the $(\ell,m)=(4,4)$ mode. 
     The time-integrated $h^{\psi_4}$ shows significant non-linearities which result in strong amplitude oscillations. The strain $h$ obtained via \eqref{e-Jtrhtl}, on the other hand, is
     essentially free of this effect.}
\label{fig:BBH-GW-signal}
\end{figure}

\begin{figure}
  \includegraphics[width=0.49\linewidth]{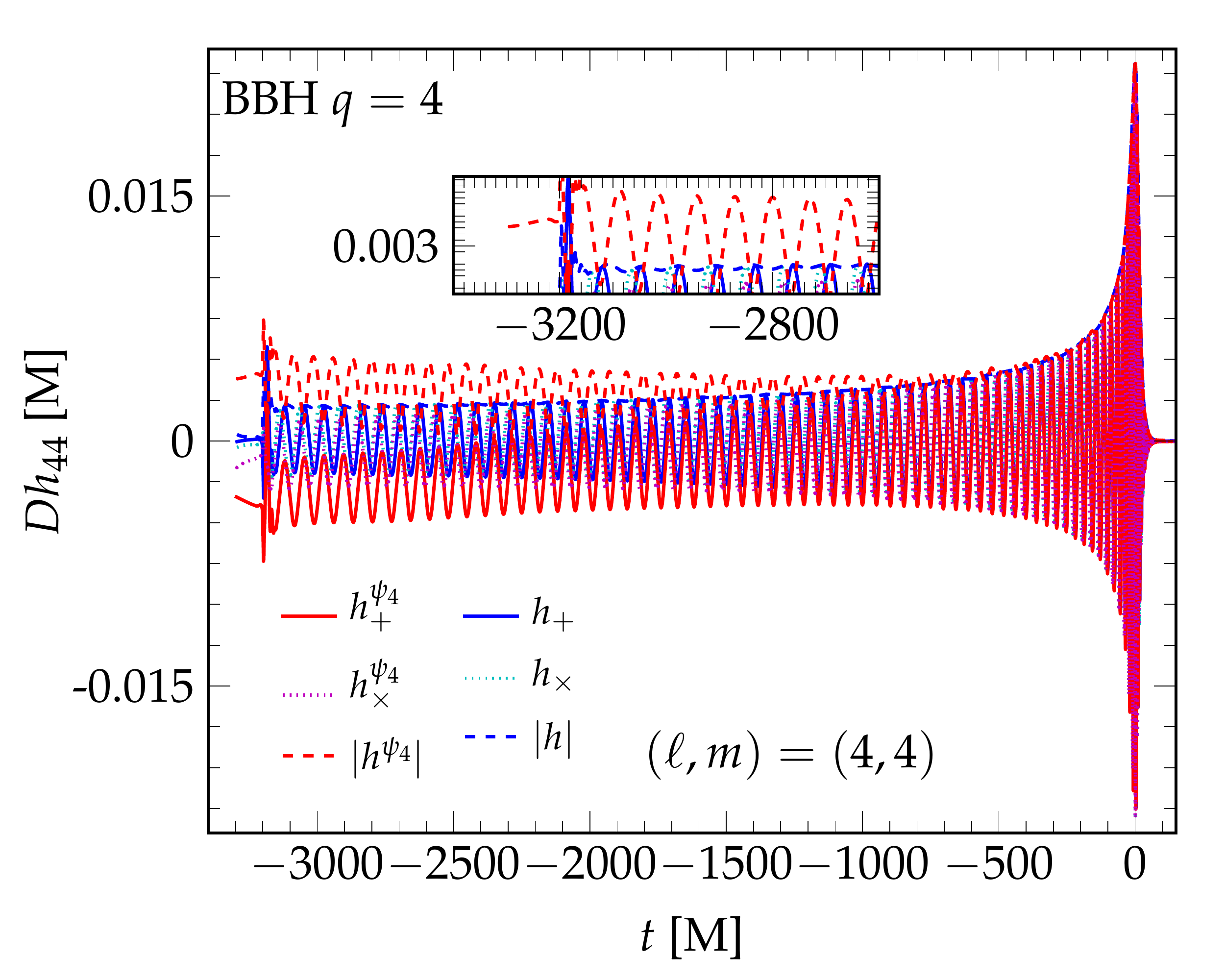}
  \includegraphics[width=0.49\linewidth]{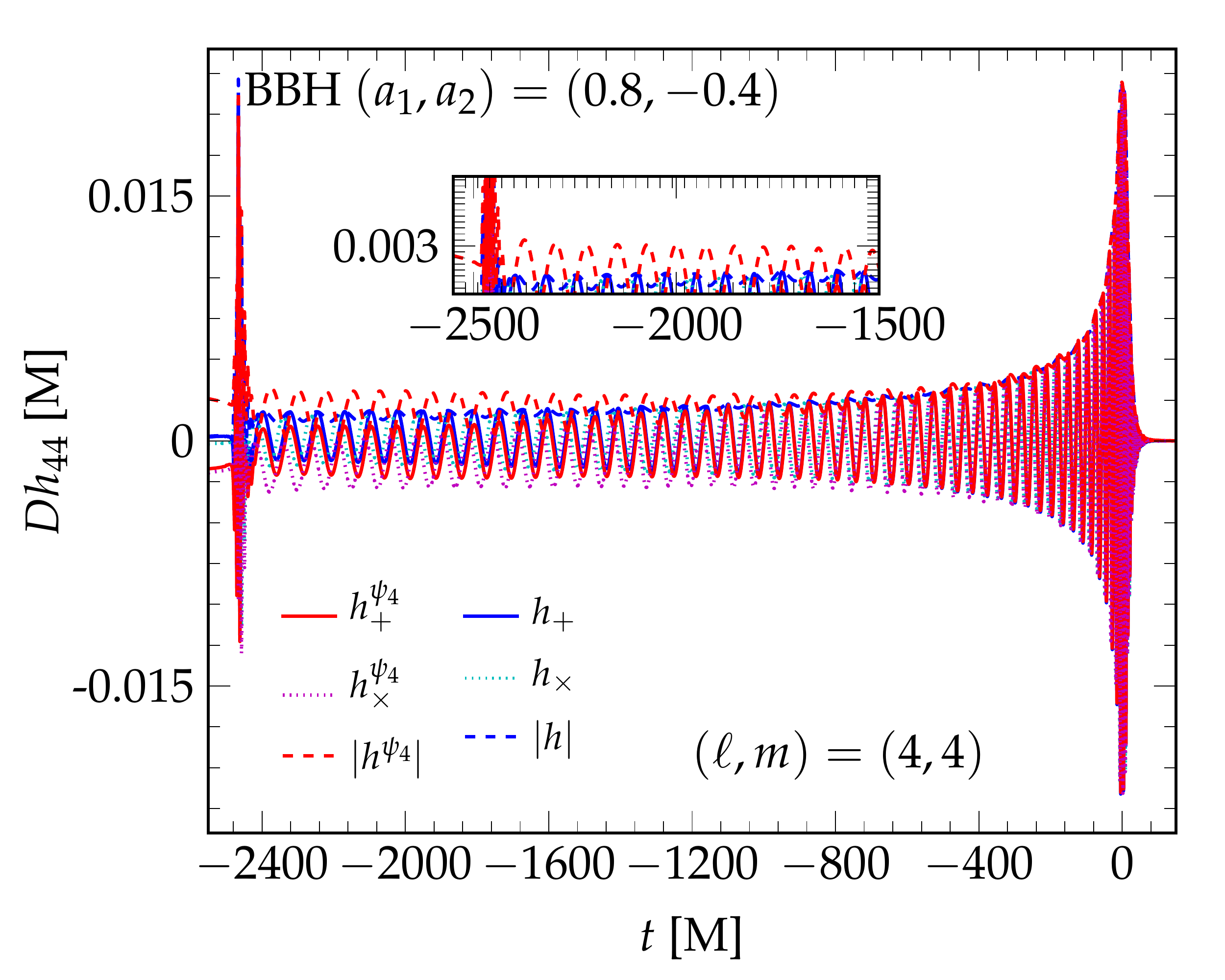}
  \caption{The same as the right panel of Fig.~\ref{fig:BBH-GW-signal}, but for a BBH system with mass-ratio $M_1/M_2=4$ (left), and an equal-mass spinning configuration $(a_1,a_2)=(0.8,-0.4)$.
           The time-integrated $h^{\psi_4}$ shows significant non-linearities which result in strong amplitude oscillations. For the strain $h$ obtained via \eqref{e-Jtrhtl}, on the other hand, 
           this effect is strongly reduced.}
\label{fig:BBH2-GW-signal}
\end{figure}

\begin{figure}
  \includegraphics[width=0.49\linewidth]{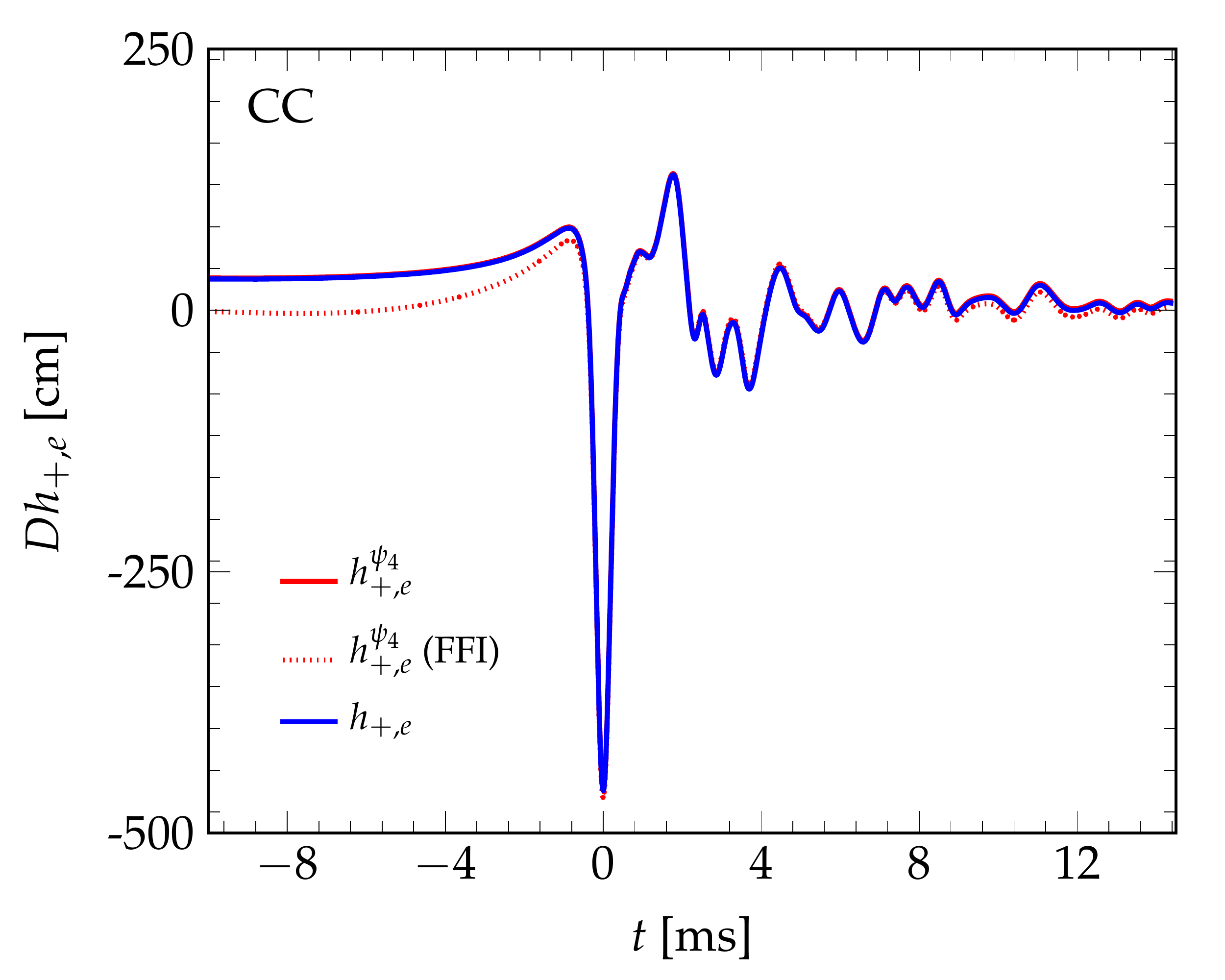}
  \includegraphics[width=0.49\linewidth]{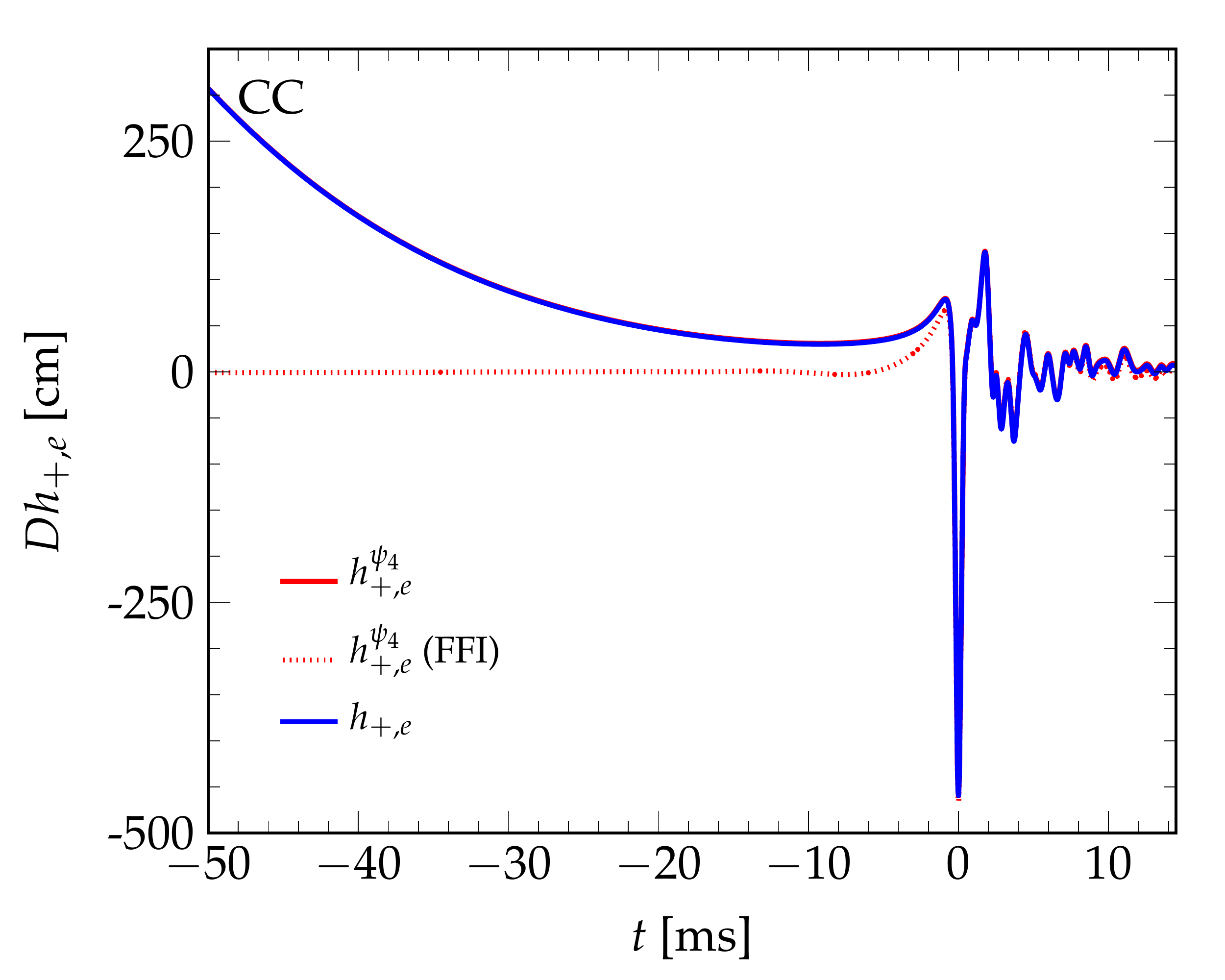}
   \caption{\textit{Left}: the distance rescaled ``+'' polarization of the gravitational wave strain $h$ as emitted in the equatorial plane from 
            stellar core collapse model ``A3B3G3''. The strain is (i) computed via \eqref{e-Jtrhtl}, (ii) integrated in the time domain from $\psi_4$, and (iii) time-integrated 
            in the Fourier domain from $\psi_4$ using FFI. \textit{Right}: The same as in the left panel, but including the entire early collapse phase. 
            Both, $h_{+,e}$ and $h_{+,e}^{\psi_4}$, should exhibit a strictly positive monotonic growth during th early collapse phase.
            Instead, both waveforms show a clear non-linear drift that leads to a significant offset.}
\label{fig:CC-GW-signal}
\end{figure}

We test the new strain formula \eqref{e-Jtrhtl} in real-world applications of Cauchy-characteristic extraction (CCE) \cite{Winicour05,Reisswig:2009us,Reisswig:2010a}.
In CCE, the strong field region of a spacetime is evolved via standard 3+1 Cauchy evolutions. Here, we use the 3+1 code described in \cite{Pollney:2009yz, Reisswig2012b}.
During 3+1 Cauchy evolution, at some finite coordinate radius defining the worldtube $\Gamma$, metric data
is collected and used as inner boundary data for a subsequent characteristic evolution to integrate the Einstein equations out to $\scri$ where 
the gravitational wave signal can be unambiguously extracted.

We use the strain formula \eqref{e-Jtrhtl} to extract gravitational radiation
at $\scri$ via CCE from simulations of three different binary black hole (BBH) merger configurations and a rapidly rotating stellar core collapse model. 
For the BBH merger test, we consider an equal-mass non-spinning black hole system with an initial separation of $D=11M$ that completes $\sim8$ orbits before merger.
The simulation, the initial parameters, and the grid setups are described in detail in \cite{Pollney:2009yz, Reisswig:2009rx}. We also consider a mass-ratio $M_1/M_2=4$ 
non-spinning configuration with an initial separation of $D=12M$ that completes $\sim 15$ orbits before merger. The simulation is described in \cite{Damour:2011fu}.
Finally, we consider an equal-mass spinning configuration with individual black hole spins that are (anti-)aligned with the orbital angular momentum.
The $z$-component of the black hole spins is given by $(a_1,a_2)=(0.8,-0.4)$. The initial separation is $D=12.5M$, and the binary completes $\sim 12$ orbits before merger.
The simulation is described in \cite{Hinder:2013oqa}.
In all cases, characteristic metric boundary data is collected
at worldtubes $\Gamma$ of radii $R_\Gamma=100M$ and $R_\Gamma=250M$ for subsequent characteristic wave extraction at $\scri$.

The rotating stellar core collapse model is taken from \cite{dimmelmeier:02}, and is labeled by ``A3B3G3''. This particular model consists of
a rapidly differentially spinning iron core at the onset of collapse, and produces a pronounced core bounce signal followed by strong 
fundamental mode excitation in the nascent protoneutron star. The initial stellar model is discussed in detail in \cite{dimmelmeier:02, Reisswig2012b}.
The simulation and grid setup are detailed in \cite{Reisswig2012b}, and a comparison between different wave extraction methods for this model has been performed in \cite{Reisswig:2011a}.
Characteristic metric boundary data is collected on worldtubes of radii $R_\Gamma=1000M$, $R_\Gamma=1500M$, $R_\Gamma=2500M$, and $R_\Gamma=3500M$.

We compare the strain as extracted via the new formula (denoted by $h$) with the strain as time-integrated from the Weyl scalar $\psi_4$ \cite{Alcubierre:2008, Babiuc:2009}
in the time domain (denoted by $h^{\psi_4}$) and in the frequency domain using
fixed-frequency integration (FFI) \cite{Reisswig:2011} (denoted by ``$h^{\psi_4}$ (FFI)'').
In Fig.~\ref{fig:BBH-GW-signal}, we show the ``$+$'' and ``$\times$'' polarizations, as well as the amplitude of $h$ and $h^{\psi_4}$
for the equal-mass binary black hole merger configuration using worldtube data from $R_\Gamma=100M$ (similar results also hold for $R_\Gamma=250M$).
We see that $h^{\psi_4}$ as time-integrated from $\psi_4$ is subject to non-linear drifts. 
This is particularly visible in the higher-order $(\ell,m)=(4,4)$ mode (right panel of Fig.~\ref{fig:BBH-GW-signal}; { Fig.~\ref{fig:BBH2-GW-signal}}), where the time-integrated $h^{\psi_4}_+$ 
does not oscillate about zero. Note that we have picked the two integration constants such that $h^{\psi_4}=0$ at late times after ring-down.
This artificial non-linear drift is also noticeable in the amplitude $|h^{\psi_4}|$ where it leads to long-lasting transient oscillations 
(see inset of left and right panel of Fig.~\ref{fig:BBH-GW-signal} { and inset of left and right panel of Fig.~\ref{fig:BBH2-GW-signal}}).
The strain $h$ as directly obtained via formula \eqref{e-Jtrhtl}, on the other hand, shows significantly better behavior for the $(\ell,m)=(4,4)$ mode, 
where essentially no artificial non-linear drifts are visible, and the amplitude $|h|$ remains monotonic (right panel of Fig.~\ref{fig:BBH-GW-signal}, 
{ and left and right panel of Fig.~\ref{fig:BBH2-GW-signal}}).
In the case of the $(\ell,m)=(2,2)$ mode { for the equal-mass non-spinning configuration}, 
there are some amplitude oscillations of $|h|$ in the initial $-1400M \lesssim t \lesssim 1000M$ due to non-linear drift. 
At around $t\simeq-1000M$, it becomes quickly monotonic, however, whereas the time-integrated $|h^{\psi_4}|$ 
is still subject to long lasting amplitude oscillations (inset of left panel of Fig.~\ref{fig:BBH-GW-signal}).
Similar behavior also holds for the other two BBH configurations (not shown). 
{ We note that the overall behavior, especially in the higher-order modes, is independent of the particular BBH merger configuration considered 
(compare right panel of Fig.~\ref{fig:BBH-GW-signal} with Fig.~\ref{fig:BBH2-GW-signal}).}

{For the equal-mass non-spinning case, we quantify the differences in $h$ as obtained via the various methods more precisely in the following.} 
When the two waveforms are aligned at merger,
we measure a relative amplitude difference between $h$ and $h^{\psi_4}$ of $\Delta A/A \simeq 0.004\%$ for the $(\ell,m)=(2,2)$ mode,
and $\Delta A/A \simeq 0.5\%$ for the $(\ell,m)=(4,4)$ mode at the peak of the signal ($t=0$).
To avoid the artifacts associated with time-integrated waveforms, 
we also directly compare $\psi_4$ to get a different handle on the 
differences between formula \eqref{e-Jtrhtl}
and $\psi_4$ extraction.
Therefore, we take the second time derivative of $h$ obtained via formula \eqref{e-Jtrhtl}. 
We arrive at relative peak amplitude differences between $\ddot{h}$ and $\psi_4$
of $\Delta A/A\simeq 0.003\%$ for the $(\ell,m)=(2,2)$ mode,
and $\Delta A/A \simeq 0.2\%$ for the $(\ell,m)=(4,4)$ mode.
To measure the phase difference, we align the signals during inspiral in the interval $t\in[-1100,800]$ using the method presented in \cite{Boyle:2008ge}, and measure the dephasing
accumulated at the merger peak. Between $h$ and $h^{\psi_4}$, we measure a relative dephasing of $\Delta\phi/\phi\simeq 0.2\%$ for the $(\ell,m)=(2,2)$ mode. 
For the $(\ell,m)=(4,4)$ mode, no meaningful dephasing can be obtained due to the large non-linear drift.
Between $\ddot{h}$ and $\psi_4$, we measure a relative dephasing of $\Delta\phi/\phi\simeq 0.0002\%$ for the $(\ell,m)=(2,2)$ mode, and 
$\Delta\phi/\phi\simeq 0.001\%$ for the $(\ell,m)=(4,4)$ mode.
The results based on the differences between $\ddot{h}$ and $\psi_4$ indicate that the new formula is 
accurate within the numerical and systematic errors inherent in CCE \cite{Taylor:2013zia}, at least for the particular case considered here.
{Similar results also hold for the unequal-mass configuration and for the spinning configuration.}

{ We now turn to the test case of rotating stellar core collapse.}
In Fig.~\ref{fig:CC-GW-signal}, we show the ``+'' polarization of the strains $h$, $h^{\psi_4}$, and ``$h^{\psi_4}$ (FFI)'' as emitted 
in the equatorial plane for the rotating stellar core collapse model ``A3B3G3''.
Instead of the expected monotonically rising signal, both, $h_+$ and $h_+^{\psi_4}$, exhibit
a noticeable non-linear drift in the first $\sim40\ \rm{ms}$ during collapse, which leads to a significant offset. 
Otherwise, $h_+$ and $h_+^{\psi_4}$ are practically identical.
Note that we have shifted the waveforms such that they oscillate about zero at late times.
The strain ``$h^{\psi_4}$ (FFI)'' does not show non-linear drifts due to the filtered low-frequency components.
As detailed in \cite{Reisswig:2011a}, this filter approach also removes frequency components that are physical. 
Unfortunately, it is not possible to disentangle the artificial low frequency components from the physical ones.
Given that directly extracting $h$ by using formula \eqref{e-Jtrhtl} does not reduce the unphysical non-linear drifts in the particular 
core collapse model that we consider, we conclude that this drift must have different roots, perhaps stemming from artifacts in the Cauchy initial data.
This is further supported by the fact that the drift is practically independent of numerical Cauchy and characteristic resolutions.
In addition, by changing the characteristic initial data from conformally flat $J=0$ to data which vanishes at $\scri$ and smoothly blends to Cauchy data via some polynomial
function, we practically observe no difference in the behavior of the drift.
A detailed analysis of this must be left to future work.

\section{Conclusion}
\label{s-con}

This work re-investigated formulas describing gravitational radiation in the characteristic
formulation of numerical relativity.
A new formula for the gravitational wave strain, $(h_+,h_\times)$, was derived.
Further, alternative procedures, that reduce the use of time integrals, were found for
calculating intermediate variables needed for any description of the gravitational radiation.
As in previous work, it is still necessary to solve the evolution equations Eq.~(\ref{e-dx0})
for $x_0^A$ and Eq.~(\ref{e-du0}) for $u_0$, but the introduction of explicit formulas for
$\omega$ (Eq.~(\ref{e-exom})) and $\delta$ (Eq.~(\ref{e-delta})) reduces the number of time
integrals from 4 to 2 in the computation of ${\mathcal N}$ and $\psi_4$, and from 5 to 2
in the computation of the wave strain.

The numerical tests show that formula \eqref{e-Jtrhtl} for directly extracting the strain $h$ at $\scri$ yields results 
which are comparable to the ``traditional'' strain computed via a double time-integration from the Weyl scalar $\psi_4$.
{ In the case of three representative binary black hole merger simulations including an equal-mass non-spinning case, a mass-ratio $M_1/M_2=4$ configuration, and
an equal-mass spinning case with spins $(a_1,a_2)=(0.8,-0.4)$}, the new formula leads to reduced artificial
non-linear drifts, { especially in higher-order modes}, which are typically observed in numerical simulations \cite{Reisswig:2011, Baker:2002qf, Berti:2007snb}.
In the case of a rapidly rotating stellar core collapse model, however, artificial drifts persist, and are thus unrelated to time integration.

The new formula has been implemented in the PITT Nullcode and is available as part of the Einstein Toolkit \cite{Moesta:2013dna, Loffler:2011ay}. 

\section*{Acknowledgments}
We would like to thank: the National Research Foundation, South Africa, for financial
support; California Institute of Technology, USA, and Max-Planck Institute for
Gravitational Physics, Germany, for hospitality; and Denis Pollney and Jeffrey
Winicour for discussions.
CR acknowledges support by NASA through
Einstein Postdoctoral Fellowship grant number PF2-130099 awarded by
the Chandra X-ray center, which is operated by the Smithsonian
Astrophysical Observatory for NASA under contract NAS8-03060.

\appendix

\section{Computer algebra scripts and output}
The Maple scripts used to derive some of the results in this paper are included in 
the online supplement. The script files are \texttt{J\_rho.map} with output
\texttt{J\_rho.out} for Eqs.~(\ref{e-Jtrht}) and (\ref{e-JtrhtLin}), and
\texttt{delta.map} with output \texttt{delta.out} for Eq.~(\ref{e-delta});
both these script files start by running the script file \texttt{common.map}.
The file \texttt{READ.ME} contains a description of the variable names used in
the scripts in terms of the notation used in this paper.

\bibliography{aeireferences,extra}

\end{document}